\documentclass[aps,preprint,superscriptaddress,showpacs,10pt]{revtex4}
\usepackage{mathrsfs}
\usepackage{amsfonts}
\usepackage{amsmath}
\usepackage{amssymb}
\usepackage{graphicx}
\usepackage{subfigure}
\usepackage{footmisc}
\usepackage{braket}
\usepackage[CJKbookmarks=true]{hyperref}

\begin{document}

\title{High fidelity quantum cloning of two known nonorthogonal quantum states via weak measurement}
\author{Ming-Hao Wang}
\affiliation{State Key Laboratory of Magnetic Resonances and Atomic and Molecular Physics,
Wuhan Institute of Physics and Mathematics, Chinese Academy of Sciences, Wuhan 430071, China}
\affiliation{University of Chinese Academy of Sciences, Beijing 100049, China}
\author{Qing-Yu Cai}
\thanks{Corresponding author. \\Electronic address: qycai@wipm.ac.cn.}
\affiliation{State Key Laboratory of Magnetic Resonances and Atomic and Molecular Physics, Wuhan Institute of Physics and Mathematics, Chinese Academy of Sciences, Wuhan 430071, China}
\begin{abstract}
We propose a scheme to enhance the fidelity of symmetric quantum cloning machine using a weak measurement. By adjusting the intensity of weak measurement parameter $p$, we obtain the copies with different optimal fidelity. Choosing proper value of $p$, we can obtain the perfect copies for initial qubits. In this paper, we focus on $1-2$ quantum cloning for two nonorthogonal states. Sets containing more than two linear independent states are also discussed briefly. Due to weak measurements being probabilistic, we obtain high fidelity at risk of probability. If the weak measurement successes, we do the following operations to obtain copies with high fidelity, otherwise, the cloning process fails and we need do nothing. From this perspective, the scheme we propose is economical for saving quantum resource and time, which may be very useful in quantum information processing.
\end{abstract}
\maketitle


\section{Introduction}
In recent years the quantum information has developed very quickly. The study of quantum information processing(QIP) has been attracting much attention from various
research communities. Various schemes for logic gates, such as C-NOT, SWAP, as required for classical computer, were proposed theoretically and implemented experimentally in many systems including optical photon~\cite{Knill2001,Nagali2009,Li2013,Tischler2017,Zeuner2017}, trapped ions~\cite{Cirac1995,Monroe1997,Kielpinski2002}, cavity quantum electrodynamics~\cite{Rauschenbeutel1999,Zubairy2003,Yan2018}, and liquid state nuclear magnetic resonance~\cite{Vandersypen2001,Baugh2007,Suter2008}. There seems great promising future in quantum computer.  It's peculiar principles such as linearity, unitarity, and inseparability have been utilized to realize quantum computer~\cite{Nielsen2011}. On one hand, these principles enhance the capacity of information processing, but, on the other hand, they put on some limitations ~\cite{Pati2002}. A fundamental restriction in QIP is that an unknown quantum state cannot be copied perfectly~\cite{Wootters1982} in contrast with replicating information ubiquitously in classical world~\cite{Wang2018}. This is a consequence of linearity of quantum mechanic. This makes a qubit so distinct from a classical bit. This limitation is known as no-cloning theorem and has found its applications in quite different fields of quantum information theory, such as quantum computation and quantum cryptography~\cite{Gisin2002}.

However, if we pay some price, then approximate or even exact cloning is possible. It dose not prohibit the possibility of approximate cloning of an arbitrary state of a quantum mechanical system. Bu{\v{z}}ek and Hillery firstly presented a scheme that given a unknown qubit, tow identical output qubits as approximate as possible to input qubit are produced~\cite{Buzek1996}. After their seminal paper, quantum cloning has been extensively studied and lot of topmost achievements have been made, both theoretically and experimentally~\cite{Fan2014,Scarani2005}. This Bu{\v{z}}ek - Hillery quantum cloning machine is state-independent and known as universal quantum cloning machine(UQCM). The optimal fidelity of a cloning bound achieves $5/6$ for UQCM. Shortly after,  a new quantum cloning machine was been proposed~\cite{Brus2000,Fan2002}, called quantum phase-covariant cloning machine(QPCCM). The fidelity of QPCCM will reach to $0.854$, which is higher than UQCM. The QPCCM is significantly important in quantum cryptography as it provides the optimal eavesdropping method for a large class of attacks on quantum cryptograph protocols ~\cite{Bennett1984,Ekert1991,Ferenczi2012}. Besides deterministic quantum cloning machine, Duan and Guo~\cite{Duan1998,Duan1998a} revealed that a quantum state secretly chosen from a linearly independent states set can be probabilistically cloned with unit fidelity and they called this quantum cloning process probabilistic quantum cloning (PQC). This kind of quantum cloning is different from the deterministic quantum cloning machine, it has nonzero probability that cloning process comes to nothing. However, once we success, we obtain the perfect copies of initial qubits. The scheme we proposed in this paper is different from the PQC. The more detailed difference between them will be discussed later.

It has well been realized that the we can manipulate qubits in a better way if we pay some price. The QPCCM is one of examples. Recently, an optimal quantum cloning machine, which clones qubits of arbitrary symmetrical distribution around the Bloch vector was investigated~\cite{Bartkiewicz2010}. More generally, states in block regime, which is a simply connected region enclosed by a ¡°longitude-latitude grid¡± on the Bloch sphere, was also investigated~\cite{Kang2016}. All those quantum cloning machines are based on the maximin principle by making full use of
a priori information of amplitude and phase about the to-be-cloned qubit input set. As expected, the performance of machines is better than UQCM. In addition to the price of limiting the range of input states, other resource can also be sacrificed, such as the probability of success. PQC is one among those.

Inspired by those previous work, we propose a new scheme that combines the weak measurement and unitary transformation. Assuming that given a qubit secretly chosen from a state set $\{\ket{\psi_1},\ket{\psi_2}\}$, our task is to duplicate the given qubits. Many related works are reported~\cite{Brus1998,Duan1998a,Zhang2012}. In their schemes, they all clones the given qubits directly. Compared with their work, our work has peculiar advantage. Before complicated quantum transformation, we pretreat the given qubits. This pretreatment allows us to obtain copies with high fidelity. Whether doing further operator depends on the result of measurement, which economize quantum resource and make our scheme a economical one. This may be serviceable in QIP. The rest of this paper is organized as follows. In Sect.II, we make a brief review of weak measurement and quantum cloning machine, especially optimal quantum cloning for two nonorthogonal states. In Sect.III, we show our scheme for higher cloning fidelity in details. Finally, a concise summary is given in Sect.V.

\section{Theory}

\subsection{Weak measurement}
The projection postulate is one of the basic postulates of the standard quantum theory and it states that measurement of a variable of a quantum system irrevocably collapses the initial state to one of the eigenstates (corresponding to the measurement outcome) of the measurement operator. Once the initial state collapses due to a projection measurement on a quantum system, it can never be recovered. However, the situation is different for the case that the measurement is not sharp, i.e., non-projective measurement~\cite{Kim2009}. For weak measurements, the information extracted from the quantum system is deliberately limited, thereby keeping the measured system¡¯s state from randomly collapsing towards an eigenstate. It is possible to reverse the measurement-induced state collapse and the unsharpness of a measurement has been shown to be related to the probabilistic nature of the reversing operation which can serve as a probabilistic quantum error correction~\cite{Koashi1999}.

Consider the initial state of a qubit is a pure state $\ket{\phi}$ and the measurement operators $P_{1}$ and $P_{2}$ are orthogonal projectors whose sum $P_{1}+P_{2}=I$ is identity. We introduce the operators

\begin{equation}\label{weak_measurement}
  \hat{M}_{yes}=\sqrt{p}\hat{P}_{1}+\hat{P}_{2}, \qquad \hat{M}_{no}=\sqrt{1-p}\hat{P}_{1}, \quad p\in [0,1].
\end{equation}
It should be noted that $M_{yes}^{\dag}M_{yes}+M_{no}^{\dag}M_{no}=I$ and therefore $M_{yes}$ and $M_{no}$ describe a measurement. Consider the effect of the operators $M$ on a pure state $\ket{\phi}$. The state can be rewritten as $\ket{\phi}=\sqrt{ p_1}\ket{\phi_1}+\sqrt{p_2}\ket{\phi_2}$, where $\ket{\phi_{1,2}}=P_{1,2}\ket{\phi}/\sqrt{p_{1,2}}$ are the two possible outcomes of the projective measurement and $p_{1,2}=\bra{\phi}P_{1,2}\ket{\phi}$ are the corresponding probabilities. The operator  $M_{yes}$ decreases the ratio $\frac{p_1}{p_2}$, causing  $\hat{M}_{yes}\ket{\phi}$ moving toward $\ket{\phi_2}$ while operator $\hat{M}_{no}$ collapses the $\ket{\phi}$ into $\ket{\phi_2}$.

\subsection{Optimal quantum cloning for two nonorthogonal states}

Suppose we are given with equal probability one quantum state from a set including two known nonorthogonal quantum states in the form
\begin{equation}\label{state_set}
  \ket{\psi_1}=\cos(\xi)\ket{0}+\sin(\xi)\ket{1}, \quad  \ket{\psi_2}=\sin(\xi)\ket{0}+\cos(\xi)\ket{1},
\end{equation}
where $\xi \in [0,\pi/4]$, with the scale product
\begin{equation}
  \bra{\psi_1}\ket{\psi_2}=\sin(2\xi).
\end{equation}
The transformation of symmetric 1-2 state-dependent cloning take the following form:
\begin{equation}\label{transformation}
  \begin{split}
    \ket{00} & \rightarrow a\ket{00}+b(\ket{01}+\ket{10})+c\ket{11},\\
    \ket{10} & \rightarrow a\ket{11}+b(\ket{10}+\ket{01})+c\ket{00},
  \end{split}
\end{equation}
where we assume the cloning coefficients a, b and c are real numbers. Due to the unitarity of the transform, the following formulas must be satisfied:
\begin{equation}\label{unitary_condition}
  a^2+2b^2+c^2=1,\quad a c+ b^2=0.
\end{equation}
Solving the Eq.(\ref{unitary_condition}), we obtain the following equations:
\begin{equation}\label{value_of_a_and_c}
  a=\frac{1}{2}(\sqrt{1-4b^2}+1),\quad c=\frac{1}{2}(\sqrt{1-4b^2}-1).
\end{equation}
In previous work, fidelity is often used as a factor of merit, which is defined as $F=\bra{\phi_{in}}\rho_{out}\ket{\phi_{in}}$, where $\rho_{out}$ is a reduced density matrix of output state $1$ or $2$. Due to the symmetry of transformation given by Eq.(\ref{transformation}), we obtain the copies fidelity as
\begin{equation}
   F(\ket{\psi_{1}})=F(\ket{\psi_{2}})=\frac{1}{4} (3 a^2+4 (a+b) (b+c) \sin (2 \xi )+(a+c) \cos (4 \xi ) (a-2 b-c)+2 a b+4 b^2+2 b c+c^2)
\end{equation}
With some calculation and using the method of Lagrange multipliers, we can determine the cloning coefficient b as
\begin{equation}\label{value_of_ b}
  b=\frac{1}{8} (1-\csc (2 \xi )+\csc (2 \xi )\sqrt{9 \sin ^2(2 \xi )-2 \sin (2 \xi )+1} ).
\end{equation}
Combine Eq.(\ref{value_of_ b}) and Eq.(\ref{value_of_a_and_c}), we obtain the detailed transformation and the maximum fidelity for optimal cloning.

\section{Scheme for higher fidelity}
In this section, we are going to investigate how to enhance the fidelity by using weak measurement. Unlike the schemes proposed by others, we do a weak measurement as a pretreatment before transformation. After measurement, we obtain new intermediate qubits. Putting them into a quantum machine, we will obtain the final qubits we want with high fidelity at the output of machine. Let us now present our scheme in detailed. Suppose we are given a qubit selected randomly from Eq.(\ref{state_set}). We first do a weak measurement described as Eq.(\ref{weak_measurement}) on the given state. Let $\hat{P}_1=\ket{+}\bra{+}$ and $\hat{P}_2=\ket{-}\bra{-}$, and $\ket{\psi_{1,2}}$ can be rewritten as
\begin{equation*}
  \ket{\psi_{1,2}}=\frac{\sqrt{2}}{2}\big(\cos(\xi)+\sin(\xi)\big)\ket{+}\pm \frac{\sqrt{2}}{2}\big(\cos(\xi)-\sin(\xi)\big)\ket{-}.
\end{equation*}
After weak measurement, if the outcome is 'yes', the initial qubits become
\begin{equation}
\ket{\psi_{1,2}'}=\frac{\hat{M}_{yes} \ket{\psi_{1,2}}}{\sqrt{p_{yes}}}=\frac{\sqrt{p} (\sin (\xi )+\cos (\xi ))}{\sqrt{(p-1) \sin (2 \xi )+p+1}}\ket{+}\pm \frac{\cos (\xi )-\sin (\xi )}{\sqrt{(p-1) \sin (2 \xi )+p+1}}\ket{-},
\end{equation}
with the probability
\begin{equation}\label{probability}
  p_{yes}=\frac{1}{2} ((p-1) \sin (2 \xi )+p+1).
\end{equation}
Similarly, we obtain the inter product of two possible intermediate states as
\begin{equation}\label{inner_product}
  \bra{\psi_{1}'}\ket{\psi_{2}'}=\frac{(p+1) \sin (2 \xi )+p-1}{(p-1) \sin (2 \xi )+p+1},
\end{equation}

Next, we do a transformation on the intermediate qubit $\ket{\psi'_{1,2}}$ and ancillary qubit, which is originally in blank state $\ket{0}$. For the sake of convenience, we substitute $\frac{\sqrt{p} (\sin (\xi )+\cos (\xi ))}{\sqrt{(p-1) \sin (2 \xi )+p+1}}$ and $\frac{\cos (\xi )-\sin (\xi )}{\sqrt{(p-1) \sin (2 \xi )+p+1}}$ with $\frac{\sqrt{2}}{2}\big(\cos(\xi')+\sin(\xi')\big)$ and $\frac{\sqrt{2}}{2}\big(\cos(\xi')-\sin(\xi')\big)$
 respectively,  so the intermediate qubits can be rewritten as£»
 \begin{gather}\label{intermediate_states}
  \ket{\psi_{1}'}=\frac{\sqrt{2}}{2}\big(\cos(\xi')+\sin(\xi')\big)\ket{+}+\frac{\sqrt{2}}{2}\big(\cos(\xi')-\sin(\xi')\big)\ket{-}=\cos(\xi')\ket{0}+\sin(\xi')\ket{1},\\
  \ket{\psi_{2}'}=\frac{\sqrt{2}}{2}\big(\cos(\xi')+\sin(\xi')\big)\ket{+}-\frac{\sqrt{2}}{2}\big(\cos(\xi')-\sin(\xi')\big)\ket{-}=\sin(\xi')\ket{0}+\cos(\xi')\ket{1},
 \end{gather}
Combining them with Eq.(\ref{inner_product}), we obtain the important equation as

\begin{equation}\label{equation}
\sin(2\xi')=\frac{(p+1) \sin (2 \xi )+p-1}{(p-1) \sin (2 \xi )+p+1}.
\end{equation}
By applying the transformation \ref{transformation}, we obtain the copy fidelity as
\begin{equation}\label{fidelity}
  F(\ket{\psi_{1}'})=F(\ket{\psi_{2}'})=\frac{1}{2} \big(1+(a+c) \cos (2 \xi ) \cos (2 \xi')+2 b (a+c)\sin (2 \xi)(\sin (2 \xi' )+1)\big).
\end{equation}
We want to derive the optimal fidelity, with the constraints as Eq.(\ref{unitary_condition}). After some algebra, we determine the clining coefficients as
\begin{gather}
  b=\frac{\csc(2 \xi) \sqrt{8 (\sin (2 \xi' )+1)^2 \sin ^2(2 \xi)+\cos ^2(2 \xi' ) \cos ^2(2 \xi)}}{8 (\sin (2 \xi' )+1)}-\frac{\cos (2 \xi' ) \cot(2 \xi)}{8 (\sin (2 \xi' )+1)}.
\end{gather}
So the optimal fidelity has an expression
\begin{equation}\label{final_fidelity}
\begin{split}
 F=&\frac{1}{32} \{16+\frac{3 \sqrt{2} \cos (2 \xi' ) \cos (2 \xi)}{(\sin (2 \xi' )+1)}\big[4 \sin ^2(2 \xi' )+8 \sin (2 \xi' )-\cos ^2(2 \xi' ) \cot ^2(2 \xi)\\
 &+\cos (2 \xi' ) \cot (2 \xi) \sqrt{ (\cos ^2(2 \xi' ) \cot ^2(2 \xi)+8 (1+\sin (2\xi' ))^2)}+4\big]^{1/2}\\
  +&\frac{\sqrt{2}\sin(2 \xi)}{(\sin (2 \xi' )+1)^2}\sqrt{ (\cos ^2(2 \xi' ) \cot ^2(2 \xi)+8 (1+\sin (2\xi' ))^2)} \big[4 \sin ^2(2 \xi' )+8 \sin (2 \xi' )-\cos ^2(2 \xi' ) \cot ^2(2 \xi)\\
  &+\cos (2 \xi' ) \cot (2 \xi)\sqrt{ (\cos ^2(2 \xi' ) \cot ^2(2 \xi)+8 (1+\sin (2\xi' ))^2)}+4\big]^{1/2}\}
\end{split}
\end{equation}
\begin{figure}[htbp]
  \centering
  \includegraphics[width=12cm]{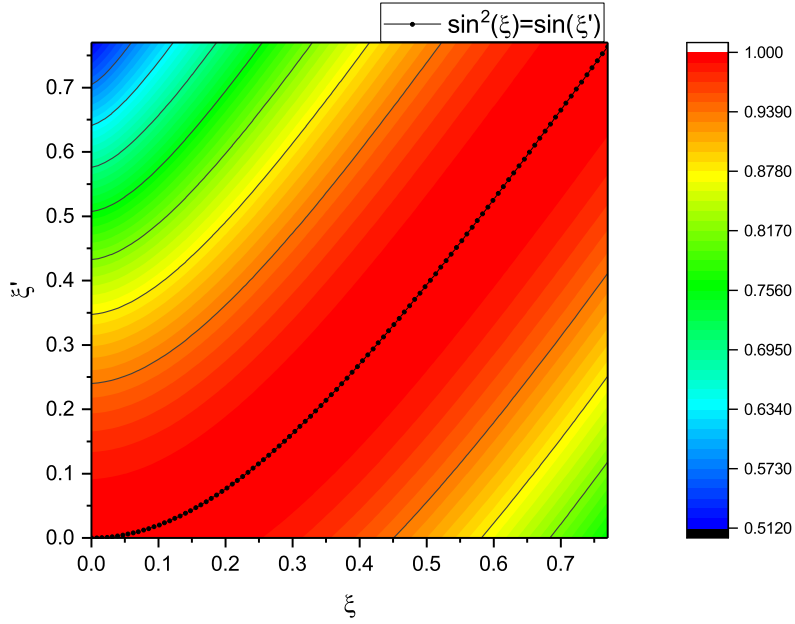}
  \caption{(Color)The dependence of the fidelity on the angle $\xi$ and $\xi'$. The dotted line correspond to $\sin(2\xi')=\sin^2(2\xi)$, in which constraint fidelity saturates unit. }
  \label{figure_fidelity}
\end{figure}

Analyzing the Eq.(\ref{final_fidelity}) as shown in Fig.(\ref{figure_fidelity}), we find that under the condition $\sin(2\xi')=\sin^2(2\xi)$, $F$ always reaches unit, which means we obtain two perfect copies of initial state. For an instant, $\xi=\pi/8$ and $\xi'=\pi/12$, which  satisfies the above condition, all parameters are determined and we obtain initial qubits
\[\ket{\psi_1}=\frac{1}{2} \sqrt{2+\sqrt{2}} \ket{0} +\frac{1}{2} \sqrt{2-\sqrt{2}} \ket{1}, \quad  \ket{\psi_2}=\frac{1}{2}\sqrt{2-\sqrt{2}} \ket{0} +\frac{1}{2} \sqrt{2+\sqrt{2}} \ket{1},\]
intermediate qubits
\[\ket{\psi'_1}=\frac{1}{4} (\sqrt{2}+\sqrt{6})\ket{0}+\frac{1}{4} (\sqrt{6}-\sqrt{2})\ket{1}, \quad  \ket{\psi'_2}=\frac{1}{4} (\sqrt{6}-\sqrt{2})\ket{0}+\frac{1}{4} (\sqrt{2}+\sqrt{6})\ket{1}, \]
transformation coefficients
\[b=\frac{1}{2 \sqrt{3}}, \quad a=\frac{1}{2} (\sqrt{\frac{2}{3}}+1), \quad c=\frac{1}{2} (\sqrt{\frac{2}{3}}-1),\]
and the final qubits
\begin{gather}\label{final_qubits}
  \ket{\psi_1}\rightarrow \ket{\psi'_1} \rightarrow  (\frac{1}{2} \sqrt{2+\sqrt{2}} \ket{0} +\frac{1}{2} \sqrt{2-\sqrt{2}} \ket{1})\otimes (\frac{1}{2} \sqrt{2+\sqrt{2}} \ket{0} +\frac{1}{2} \sqrt{2-\sqrt{2}} \ket{1},\\
  \ket{\psi_2}\rightarrow \ket{\psi'_2} \rightarrow  (\frac{1}{2} \sqrt{2-\sqrt{2}} \ket{0} +\frac{1}{2} \sqrt{2+\sqrt{2}} \ket{1})\otimes (\frac{1}{2} \sqrt{2-\sqrt{2}} \ket{0} +\frac{1}{2} \sqrt{2+\sqrt{2}} \ket{1},
\end{gather}
which are exactly the perfect copies of the initial qubits.
\begin{figure}[htbp]
  \centering
  \subfigure[$\xi=\pi/16$]{
  \label{16}
  \includegraphics[width=2.9in]{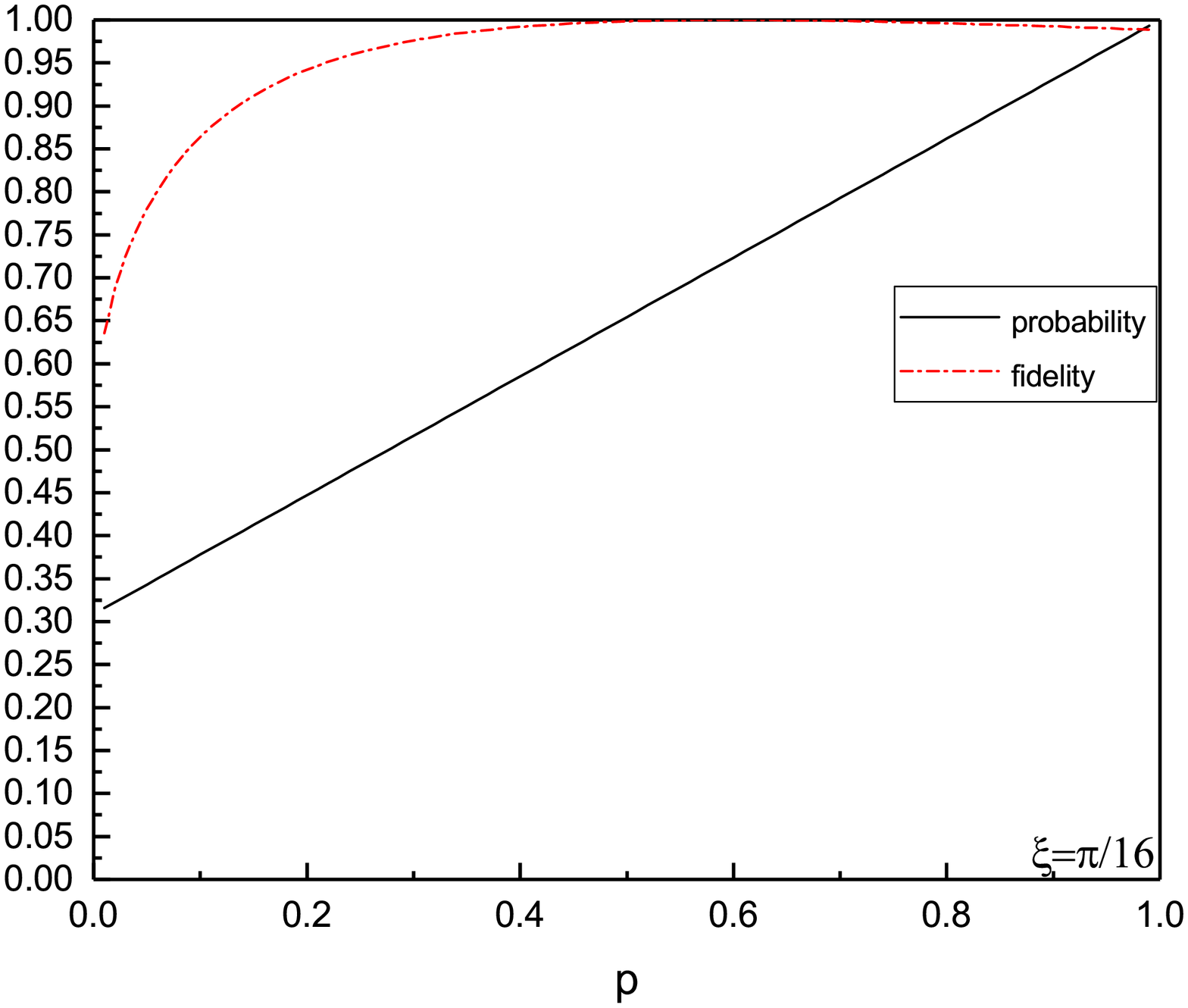}}
  \hspace{0.2in}
  \subfigure[$\xi=\pi/12$]{
  \label{12}
  \includegraphics[width=2.9in]{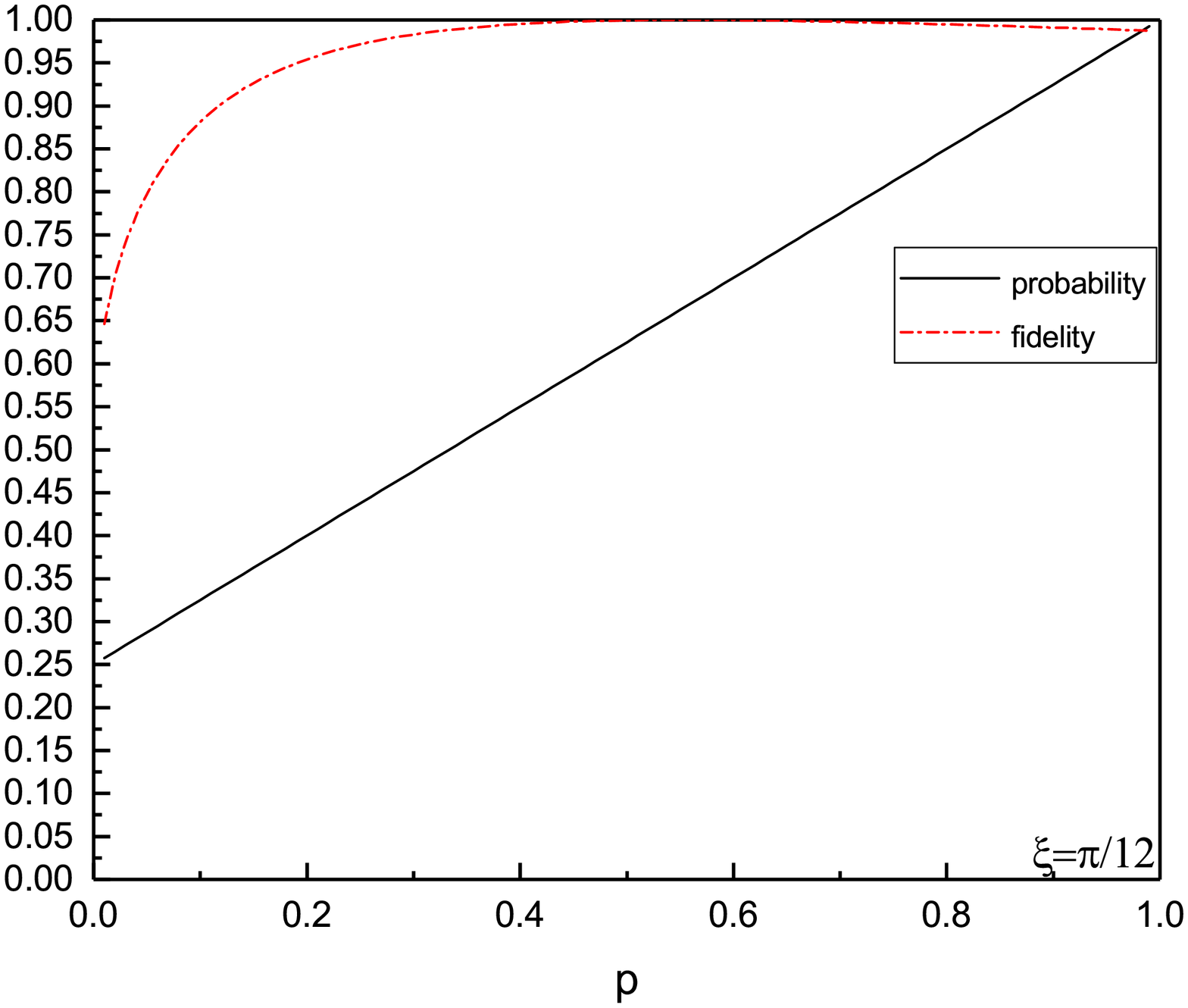}}

  \subfigure[$\xi=\pi/8$]{
  \label{8}
  \includegraphics[width=2.9in]{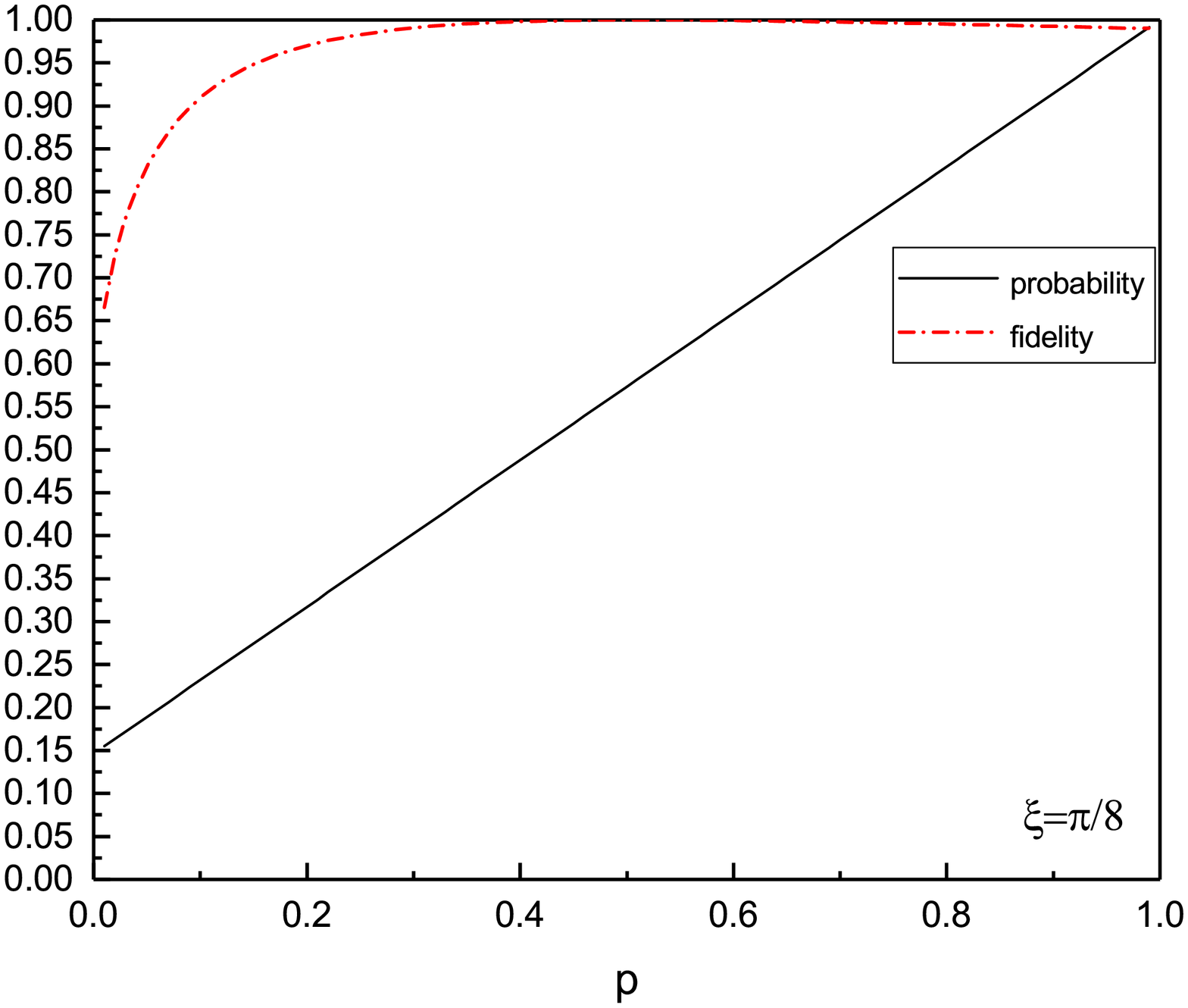}}
  \hspace{0.2in}
  \subfigure[$\xi=\pi/6$]{
  \label{6}
  \includegraphics[width=2.9in]{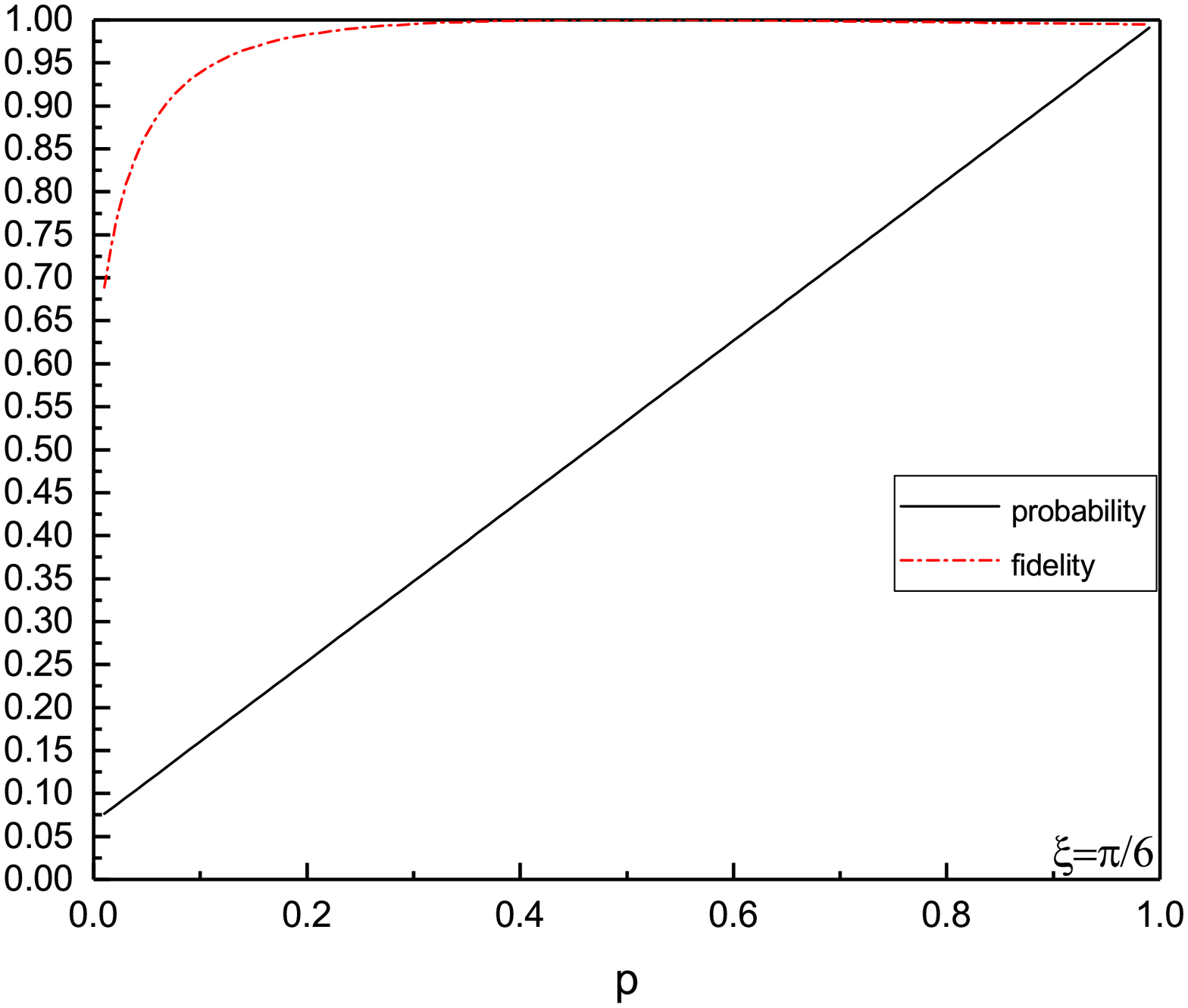}}

  \caption{(Color)The dependence of the probability and the fidelity  on the parameters of $p$ and $\xi$.}
  \label{dependence_on_p}
\end{figure}

Next, we discuss the probability of success, which have the formula as Eq.(\ref{probability}). It is function of $\xi$ and $p$, where $\xi$ arranges from $0$ to $\pi/4$ and $p$ arranges from $0$ to $1$. Generally, $\xi$ is known and unchanged. We choose proper value of $p$ so that we keep a balance between probability of success and fidelity. It has been shown dependence of probability of success and fidelity on $p$ in the Fig.(\ref{dependence_on_p}) that given a fixed $\xi$. Probability is monotone increasing while fidelity increase at first and then decrease after it reach unit.  It is worth noting that in the case of fidelity being unit, we obtain the relationship as $\frac{(p+1) \sin (2 \xi )+p-1}{(p-1) \sin (2 \xi )+p+1}=\sin^2(2\xi)$, corresponding the probability as  $p_{yes}=\frac{1}{1+\sin(2\xi)}$, which is just the Duan-Guo bound~\cite{Duan1998a}.
\section{conclusion}
In this paper, we propose a scheme to duplicate qubits chosen randomly from a nonorthogonal state set using weak measurement. We have demonstrated that weak measurement can indeed be useful for high fidelity of quantum cloning process. Compared to general Quantum cloning machine, we firstly do a weak measurement on the qubit to make it easier to clone. If the result of a measurement is 'yes', we feed the qubit to quantum cloning machine. After cloning process accomplished, we obtain copies with high fidelity dependent on the value of $p$. By choosing proper $p$, we can even obtain the perfect copies. It is easily implemented with current experimental techniques. It is worth emphasize that our scheme is also work for the state set that contains more than two linear independent states. Analogously, what we need to do is find a set of proper operators which pretreat the qubits before unitary transformation. Since the weak measurement is non-unitary, the whole process is probabilistic and the probability of success is dependent on $p$. Sometimes, we may get nothing. But, at the risk of failure, we can obtain higher fidelity of one of the copies. There are several difference between our scheme and PQC. The main difference is that for PQC, we can only obtain perfect copies with some probability or nothing, while for our scheme, we can adjust $p$ to achieve balance between probability of success and fidelity of copies. This flexibility may be of great useful in QIP. For instants, for some quantum key distribution, using the our cloner, Eve can sacrifice the success probability in order obtain information, and Alice and Bob can not detect whether they are eavesdropped if they only use the disturbance criterion. Eve hides himself in qubits loss. In general, the best eavesdropping scheme for Eve is to choose proper $p$ to keep a balance between probability and fidelity so that he obtain information as much as possible and hide oneself in the qubits loss and noise from discovering. What's more, we do a weak measurement before cloning transformation, which consist of  a series of complicated quantum logic gates in experiment. Only when the outcome is 'yes', we do the further transformation, otherwise, we quit. This can save us lots of resource, such as time and quantum qubits(ancillary qubits). Thus, our scheme is economic from this perspective. It should also be emphasized that the method we used here is not restricted to quantum cloning process. It is suitable in many other situation in QIP. The core thought is that we pretreat the target qubits, in which some price may be payed. However, we may complete the mission beyond the limitation.
\section*{Acknowledgments}

Finial support from National Natural Science Foundation of China under Grant Nos. 11725524, 61471356 and 11674089 is gratefully acknowledged.

\bibliography{database_high}

\end{document}